\documentclass[12pt,preprint]{aastex}

\newcommand{\Msun}{\mbox{$M_\sun$}}
\newcommand{\twos}{2S\,1145$-$619}
\newcommand{\onee}{1E\,1145.1$-$6141}
\begin{document}

\title{The Orbit of the High-Mass X-Ray Binary Pulsar \onee}
\author{Paul S. Ray}
\affil{E.~O. Hulburt Center for Space Research, Code 7655, 
Naval Research Laboratory, Washington, DC 20375; Paul.Ray@nrl.navy.mil}

\medskip
\and

\author{Deepto Chakrabarty\altaffilmark{1}}
\affil{Department of Physics and Center for Space Research,
Massachusetts Institute of Technology, Cambridge, MA 02139; 
deepto@space.mit.edu}

\altaffiltext{1}{Alfred P. Sloan Research Fellow}

\begin{abstract}
Observations of the 297~s X-ray pulsar \onee\ with the {\em Rossi
X-Ray Timing Explorer} have revealed its 14.4~d eccentric orbit around
its B supergiant companion.  The best-fit orbital elements are:
$P_{\rm orb} = 14.365(2)$~d, $a_x\sin i=99.4(18)$~light~s,
$e=0.20(3)$.  No eclipses are detected, indicating that the binary
inclination is $\lesssim 55^\circ$.
\end{abstract}


\section{Introduction}

It is a remarkable coincidence that there are two unrelated X-ray
pulsars in Centaurus with nearly identical spin periods separated by
only 15\arcmin\ on the sky \citep{wps78,lmh+80}.  One of the sources
is the 292~s transient X-ray pulsar \twos, which is associated with
the main sequence Be~V companion Hen~715 (HD 102567) \citep{dab+78}
and is 1.5~kpc distant.  The pulsar exhibits periodic outbursts at
186.5~d intervals, which are believed to occur during periastron
passage in the neutron star's eccentric orbit.  The X-ray flux in
quiescence is typically $\la$ 3 mcrab but the periastron flares reach
a flux of several hundred mcrab.

The second source is the 297~s X-ray pulsar \onee, which is associated
with the B2~Iae supergiant companion V830~Cen \citep{hcc81,dc82} and
is $8.5 \pm 1.5$ kpc distant.  The pulsar appears to be persistent and
steady, with a typical X-ray flux of a few mcrab, corresponding to a
luminosity of order $10^{36}$ erg~s$^{-1}$.  Such a low luminosity is
inconsistent with Roche-lobe overflow and indicates that the pulsar is
almost certainly accreting from the wind of V830~Cen.   A binary
period of $\gtrsim 6$~d is required for the B~supergiant to fit within
the pulsar's orbit \citep{dc82}, and previous authors have
proposed periods ranging from 5.6~d to 12.1~d based on optical
photometric and spectroscopic studies \citep{icm82,hcct87}.  

For supergiant X-ray binaries with orbital period $P_{\rm orb}\lesssim
20$~d, there is a significant \textit{a priori} probability of an
X-ray eclipse (corresponding to a critical binary inclination) given
approximately by
\begin{equation}
\Pr \simeq 0.38 \left(\frac{P_{\rm orb}}{\mbox{\rm 15\ d}}\right)^{-2/3}
      \left(\frac{M_c}{15 M_\odot}\right)^{-1/3}
      \left(\frac{R_c}{24 R_\odot}\right) ,
\end{equation}
where $M_c$ and $R_c$ are the companion's mass and radius, and we have
assumed a circular orbit.  Eclipsing X-ray pulsars provide important
constraints on the neutron star mass range \citep{vkvpz95}.  Since only
8 such systems are known \citep{cwj+02,cm02}, it is of significant interest to
increase the sample.  Thus motivated, we observed \onee\ with the {\em
Rossi X-Ray Timing Explorer (RXTE)} in an effort to determine the
pulsar's orbital period and search for X-ray eclipses.  In this paper,
we report our discovery of the 14.4~d orbit of \onee.  We found no
evidence for an X-ray eclipse.

\section{Observations}

Our observations were made using the Proportional Counter Array (PCA)
on {\em RXTE} \citep{jsg+96}.  This detector consists of five 
identical proportional counter units (PCUs), each containing a propane
anticoincidence layer followed by several xenon/methane layers.  It
operates in the 2--60~keV range and has an effective area of
$\sim$6500 cm$^2$ and a 1$^\circ$ field of view.  In addition to the 
standard data modes, data were also collected in GoodXenon mode, which
records the arrival time (1 $\mu$s resolution) and energy (256-channel
resolution) of every non-vetoed event.   To maximize the
signal-to-noise ratio, we confined our analysis to events 
in the top xenon layer of each PCU and restricted the energy range to
2.5--10~keV.

We observed \onee\ on multiple occasions between 1997 June and 2000
February.  Our 1997 observations were each 20~ks long, while the
1998--2000 observations were short 3~ks observations at roughly daily
intervals.  A summary of these observations is given in
Table~\ref{tab-obs}.  In all cases, the observation epochs were
selected to avoid periastron passages of \twos\ in order to prevent
contamination by bright outbursts of this source.  It is important to
note that non-imaging timing observations can only resolve the
coherent pulsations of \onee\ and \twos\ into separate Fourier bins
for observations lengths $\gtrsim 20$~ks.  Based on our long
observations where the two nearby periodicities can be distinguished,
in quiescence (i.e., away from its periastron),
\twos\ does not interfere significantly with observations of \onee.
The PCA energy spectrum for \onee\ during the various observations was
well fitted by an absorbed, cutoff power law spectral model.  Using
our 1997 June 30 observation (14.6~ks) as a typical example, the
best-fit spectral parameters were: photon index $\Gamma=1.24$, cutoff
energy $E_c=6.4$~keV, $e$-folding energy $E_f=18$~keV, and equivalent hydrogen
column density $N_{\rm H}=3.3\times 10^{22}$ cm$^{-2}$.  These values
are typical of the rest of our observations, but the source flux was
variable by a factor of $\sim$2.

\section{Timing Analysis}

\begin{figure}
\resizebox{3.0in}{!}{
\includegraphics[angle=270]{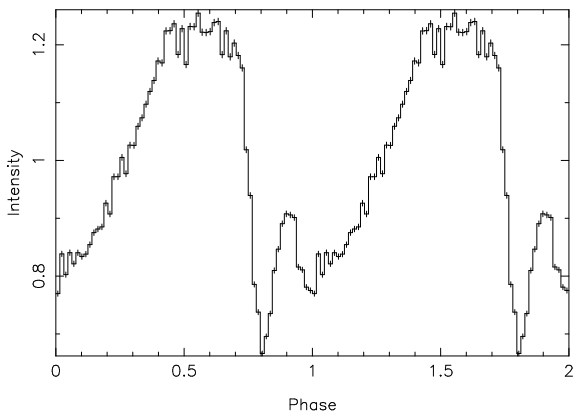}
}
\caption{Folded light curve (2--10 keV) of \onee\ with 64 bins from a
14.6 ks observation beginning 1997 June 21 02:28 UTC.  Two
cycles are shown.  Vertical scale is relative count rate.}
\label{fig-lcurve}
\end{figure}

\begin{figure*}
\resizebox{\textwidth}{!}{
\includegraphics{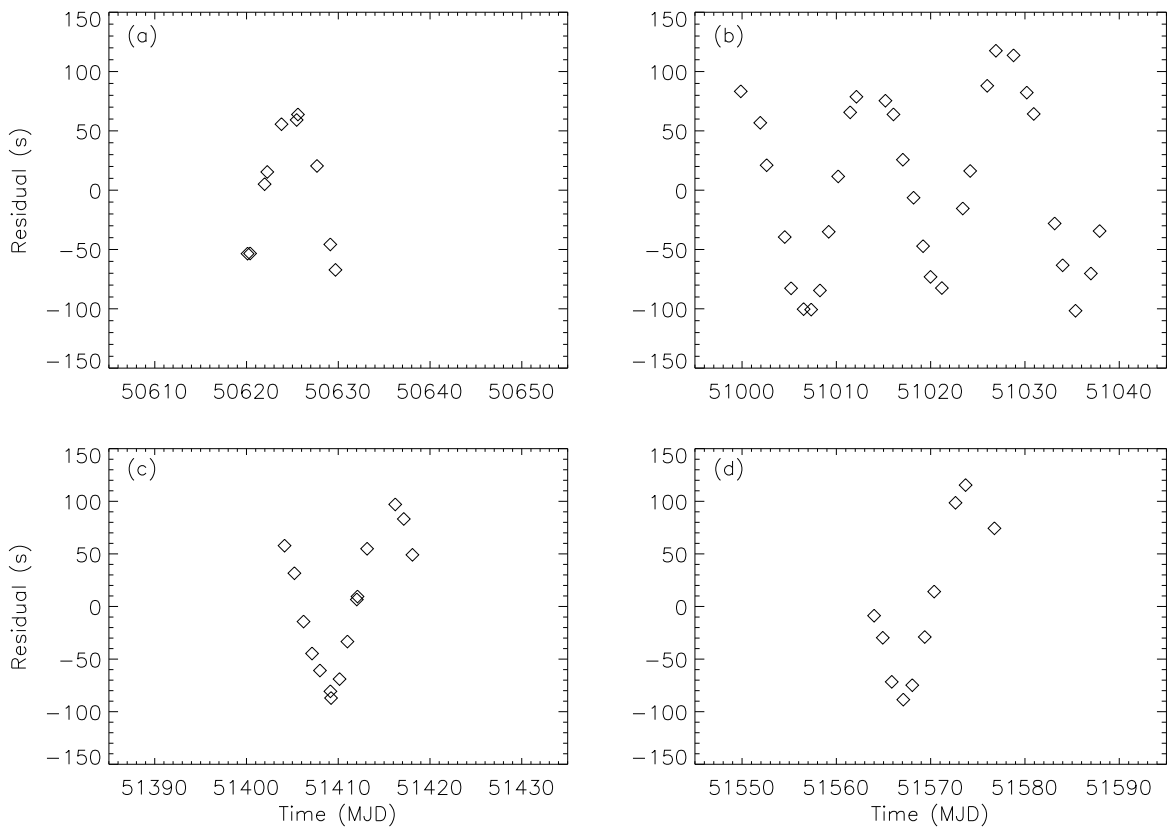}
}
\caption{Observed pulse arrival times for \onee\ from four sets of
\textit{RXTE} observations after subtracting a constant-period model.}
\label{fig-toas}
\end{figure*}

We binned the 2--10~keV events from each observation into a time
series at 1~s resolution and converted to barycentric dynamical time
(TDB) at the solar system barycenter.   We folded the time series from
each observation at a nominal pulse period of 296.5~s to form a 64-bin
pulse profile and then cross-correlated each pulse profile with a high
signal-to-noise template (see Figure~\ref{fig-lcurve}) in order to
derive a pulse arrival time for each observation.   These arrival
times can be compared to the time predicted by a constant-period model,
using the best-fit pulse period determined for each of the four
observation epochs.  The resulting arrival time residuals are shown in
Figure~\ref{fig-toas}.  There is clearly a roughly sinusoidal
variation in the residuals with a period of about 14 days in the
residuals, which we presume is due to orbital motion.   We determined
an initial orbital period using the measured minima in the arrival time
residuals in Figure~\ref{fig-toas}.  Assuming a constant orbital
period $P_{\rm orb}$, these five minima must be separated by 
integer multiples of $P_{\rm orb}$.  The best-fit orbital period is
thus $P_{\rm orb}=14.37 \pm 0.02$ d.

In order to derive the remaining binary parameters and refine the
orbital period determination, the effects of the binary motion must be
decoupled from intrinsic changes in the pulsar's spin period due to
accretion torques.  We perfomed a combined fit of all of the arrival
time measurements shown in Figure~\ref{fig-toas}.  Because of the
effect of accretion torque during the time between observations we
could not produce a fully phase-connected orbital fit.  Instead, we
used a model where the frequency and phase of the pulsar were allowed
to jump discontinuously between each epoch of observation, while the
orbital parameters applied globally.  We performed this fit using the
TEMPO pulsar timing package (see http://pulsar.princeton.edu/tempo).
No frequency derivatives within each observation epoch were included
in the model since local accretion torques seem to dominate on
a timescale of days.

\begin{figure}
\resizebox{3.0in}{!}{
\includegraphics{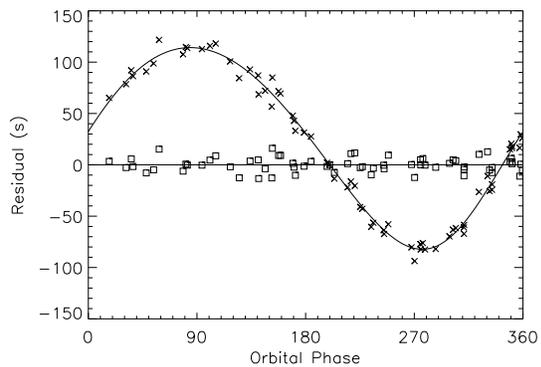}
}
\caption{Observed pulse arrival time residuals (with each epoch
allowed an arbitrary jump in pulse period and phase).
The best-fit eccentric orbit solution (from
Table~\ref{tab-orbit}) is overplotted as a solid line.}
\label{fig-orbfit}
\end{figure}

Our best-fit binary parameters are given in Table~\ref{tab-orbit}.
This fit is shown in Figure~\ref{fig-orbfit}.  The errors quoted in
Table~\ref{tab-orbit} are statistical errors only; any biases
introduced from the accretion torque are not included.  However, these
systematic errors should be reasonably small since the data cover
more than six cycles of the binary orbit and such effects can be
expected to average out.

\begin{figure}
\resizebox{3.0in}{!}{
\includegraphics{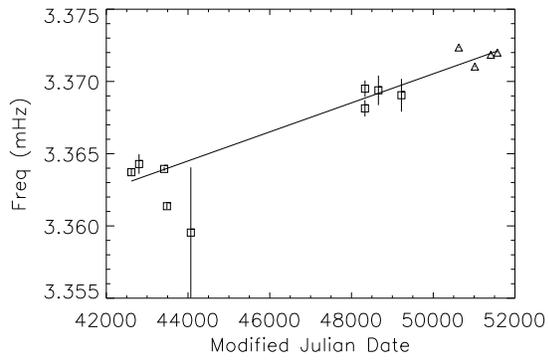}
}
\caption{Barycentric pulse frequency as a function of time for \onee.
Squares are previously published data points and triangles are from
our \textit{RXTE} observations.}
\label{fig-per}
\end{figure}

With the orbit determined, we can determine the best fit pulsar spin
frequency at each epoch with the orbital effects removed.  These
determinations are shown in Table~\ref{tab-periods}, and
Figure~\ref{fig-per}.  In each case, we have fit a constant frequency
model to determine the best average frequency for the epoch.  In all
cases there are significant deviations from this model.  The errors
shown are the single parameter standard deviations not including
coupling to the other parameters which were held constant.  For
completeness, the historical measurements of the pulsar period are
also included in the table.  However, it should be noted that they are
not corrected for the Doppler effect of the pulsar orbit and may be
inconsistently corrected to the solar system barycenter.  The pulsar
projected orbital velocity is 150 km/s so the reported periods could
differ from the intrinsic spin period by as much as 0.002 mHz.  

The pulsar has shown significant spin up since its discovery in 1978.
Fitting the frequencies to a straight line yields an average frequency
derivative of $1.2 \times 10^{-14}$ Hz/s.  This implies a spin up
timescale of $9\times 10^{3}$ y, which is well within the range of the
known supergiant X-ray pulsars. For observers, the predicted pulse
frequency of
\onee\ is
\begin{equation}
\nu(\mathrm{mHz}) = 3.36 + 1.0 \times 10^{-9} T,
\label{eq-period}
\end{equation}
where $T$ is the desired JD - 2440000.0.

\section{Discussion}

We have found that the supergiant X-ray pulsar \onee\ is in a
moderately eccentric 14.4 d orbit.  On a $P_{\rm spin}$-$P_{\rm orb}$
diagram, the system resides among the wind-fed supergiant X-ray
binaries \citep{cor86,wv89,bcc+97}, as expected from its low X-ray
luminosity.  The absence of an X-ray eclipse may be used to constrain
the binary inclination (and hence the supergiant mass) for a given
companion radius $R_c$.  Neglecting the moderate orbital eccentricity,
the no-eclipse condition is $R_c/(a_x\sin i) < \cot i$, where we have
assumed that the companion mass is much larger than the pulsar mass.
For the range of radii consistent with a B2~Ia supergiant (30--60
$R_\odot$; see de Jager \& Nieuwenhuijzen 1987)\nocite{dn87}, this
gives a maximum inclination angle ranging from 55$^\circ$ to
40$^\circ$.  Assuming a neutron star mass of 1.4 $M_\odot$, this
implies a minimum companion mass of 11--22 $M_\odot$.

\citet{prps02} have noted that massive X-ray binaries can be
divided into three broad groups: (i) moderately wide ($P_{\rm
orb}\simeq$ 20--100 d) binaries with a significant ($e>0.3$)
eccentricity caused by a neutron star ``kick'' at birth; (ii)
short-period, ($P_{\rm orb}\lesssim 10$ systems in which tidal
circularization has resulted in low-eccentricity ($e\lesssim 0.1$)
orbits;  and (iii) wide ($P_{\rm orb}> 30$ d) binaries with low
($e<0.2$) eccentricities, which may have experienced a weaker kick at
birth.  We note that \onee\ does not fit comfortably into any of these
categories, although it appears to be intermediate between groups (i)
and (ii).  As a 14.4-d binary, it is possible that tidal torques have
played some role in reducing the eccentricity caused by the neutron
star birth, although this is unlikely to have been a strong effect.
It would be interesting to  identify more binaries in the 10--20~d
range, in order to better understand at what point tidal torques play
a substantial role. 

\acknowledgements

Basic research in X-ray astronomy at NRL is funded, in part, by the
Office of Naval Research.  This work was partially supported by NASA
DPR No. S-42631-F.


\begin{deluxetable}{lcl}
\tablewidth{0pt}
\tablecaption{RXTE Observations of \onee\label{tab-obs}}
\tablehead{
\colhead{ObsId} & \colhead{\# Obs} & \colhead{Date Range}\\
}
\startdata
20130-01-* & 6\tablenotemark{a} & 1997 Jun 21--Jul 30  \\
20130-02-* & 4\tablenotemark{b} & 1997 Sep 24--Aug 16  \\
30108-01-* & 40 & 1998 Jul 4--Aug 12  \\
40079-0[1,2,3]-* & 20 & 1999 Aug 14--Aug 28  \\
40079-04-* & 14 & 2000 Jan 21--Feb 1  \\
\enddata
\tablenotetext{a}{\onee\ not detected in 20130-01-05-00}
\tablenotetext{b}{Contaminated by \twos\ flare, so not used}
\end{deluxetable}

\begin{deluxetable}{cc}
\tablewidth{0pt}
\tablecaption{Orbit Parameters
\label{tab-orbit}}
\tablehead{
\colhead{Parameter} & \colhead{Value}}
\startdata
$P_\mathrm{orb}$ & $14.365 \pm 0.002$ d \\
$T_{0}$ & $51008.1 \pm 0.4$  \\
$a_X \sin i$ & $99.4 \pm 1.8$ lt-s \\
$e$ & $0.20 \pm 0.03$ \\
$\omega$ & $-52^\circ \pm 8^\circ$ \\
$f_X(M)$ & $5.1 \pm 0.3$  \Msun \\
\enddata
\tablecomments{Quoted errors are statistical only.}
\end{deluxetable}

\begin{deluxetable}{llll}
\tablewidth{0pt}
\tablecaption{\onee\ Pulse Frequency Measurements
\label{tab-periods}}
\tablehead{
\colhead{Epoch (MJD)} & \colhead{$\nu$ (mHz)} & \colhead{Instrument} &
\colhead{Reference}
}
\startdata
42614  & 3.3637(2)\tablenotemark{a}  & \textit{OSO-8}    & \citealt{wpb+80} \\
42802  & 3.3642(7)\tablenotemark{a}  & \textit{OSO-8}    & \citealt{wpb+80}\\
43417  & 3.3639(1)\tablenotemark{a}  & \textit{Ariel 5}  & \citealt{wpb+80}\\
43484  & 3.3614(2)\tablenotemark{a}  & \textit{Ariel 5}  & \citealt{wpb+80}\\
44069  & 3.359(4)\tablenotemark{a}   & \textit{Einstein} & \citealt{wpb+80}\\
48329.2  & 3.3681(6)\tablenotemark{a}  & \textit{Ginga}  & \citealt{mih95} \\
48330.7  & 3.3695(6)\tablenotemark{a}  & \textit{Ginga}  & \citealt{mih95}\\
48658.8  & 3.369(1)\tablenotemark{a}   & \textit{Granat} & \citealt{gps92} \\
49222    & 3.369(1)\tablenotemark{a}   & \textit{ASCA}   & \citealt{smc+97} \\
\hline
50625    & 3.37235(5) & \textit{RXTE}     & this work \\
51018    & 3.37103(5) & \textit{RXTE}     & this work \\
51410    & 3.37185(2) & \textit{RXTE}     & this work \\
51570    & 3.37200(3) & \textit{RXTE}     & this work \\
\enddata
\tablenotetext{a}{Not corrected for pulsar orbital motion}
\end{deluxetable}


\end{document}